# Clustering of Content Supporting Computer Mediated Courseware Development

G. M. M. Bashir, M. J. Hossain and M. R. Karim

**Abstract**—Computer Mediated Courseware (CMC) has been developed so far for individual courses considering single or multiple text books. A group of courseware can be developed by using multiple text books and in this case, it is a requirement to cluster the contents of different books to form a generalized clustered content. No work has been found to develop courseware applying generalized clustered content. We have proposed a clustering of content supporting computer mediated courseware development based on data mining techniques to construct a hierarchical general structure of a group of courseware combining the individual structure of a set of books. The clustering will help the courseware developer to dynamically allocate contents to develop different courses using a group of books. The authors have applied this methodology for different level of courses on database. The methodology is generalized and can be applied to any other courses.

——————————————  ◆  ——————————————

## 1 INTRODUCTION

COMPUTER Mediated Courseware (CMC) is educational computer software whose primary purpose is teaching or self-learning [1]. Computer Mediated Courseware (CMC) have been developed for different level of education e.g., distance learning [2-4], Engineering Education [5] and others [6,7,8] focusing on lecture video, audio or slide representation. Existing CMC for engineering education is mainly guided by teacher centric and non-interactive. It is difficult to achieve the total coverage of the courseware. The effectiveness of a courseware depends on the organization of the courseware covering the total spectrum of the course.

Computer Mediated Courseware (CMC) has been developed so far for individual courses considering single or multiple text books. A group of courseware can be developed by using multiple text books and in this case, it is a requirement to cluster the contents of different books to form a generalized clustered content.

The basic element of a courseware is the topic that covers an atomic course item. An atomic item means that the item describes an idea or a theory or a rule that cannot be further decomposed. Multiple topics form a section/subsection. A set of sections that elaborates similar things forms a chapter. A courseware consists of a number of chapters.

In general a courseware is developed based on several text books written by different authors. Different text book organizes topics in different ways. A courseware based on different text book should have a general hierarchy of *chapters⟶sections⟶subsection⟶topics* such that it covers all the topics covered by different texts. It is a difficult job to develop this generalized hierarchy covering all the topics. Courseware so far has been developed based on a single text book and just referring the URL of other books or name and author names of books. This paper describes a generalized methodology of clustering the courseware topics to form the generalized hierarchy of a courseware. The authors have used similarity based clustering technique to form the higher level clusters of the content of a courseware based on multiple text books. The authors have considered three text books on Database Management System (DBMS), the core course on Computer Science and Engineering for the implementation of this clustering of contents. Section 2 describes the literature survey on courseware development methodology. Section 3 describes the structure of CMC and general structure of group courseware. Section 4 describes transformation of book tree into relational representation. Section 5 elaborates the clustering procedure. Section 6 is the result and discussion. Section 7 is the conclusion.

## 2 LITERATURE SURVEY

The mystification surrounding the term 'online course' arises because it is used indiscriminately to apply to nearly any course which makes even a passing use of the Internet, as well as to those where every aspect of the course is only accessible electronically [2]. John Bourne et al. [5] have discussed the Sloan Consortium's quest for

————————————————

- G. M. M. Bashir is with the Department of Computer and Communication Engineering, Faculty of Computer Science and Engineering, Patuakhali Science and Technology University, Patuakhali, Bangladesh.
- M. J. Hossain is with the Department of Computer and Information Technologyt, Faculty of Computer Science and Engineering, Patuakhali Science and Technology University, Patuakhali, Bangladesh
- M. R. Karim is with the Bangladesh Bank, Bangladesh

quality, scale and breadth in online learning, the impact on both continuing education of graduate engineers as well as degree-seeking engineering students, and the future of engineering colleges and schools as worldwide providers of engineering education. The National Teaching and Learning Database (NTLD) project in Australia has been designed to provide access to learning materials [6]. Wade et al. [7] proposes an automated, third party WWW based evaluation service which focuses on usability issues of WWW based courseware and which can be used by any WWW course instructor/Student. This paper researches the design, development and trialing of a WWW based evaluation service for WWW courseware. The paper concludes with an assessment of the benefit of using such an evaluation service to improve WWW based courseware. Hoic-Bozic et al. [8] present the result of the questionnaire about the effectiveness and quality of the courseware and the level of student acceptance of courseware as a teaching resource. Galvao et al. [9] present research that provides an analysis of various courseware features that are available on the Internet, in order to develop a new model that is based on some of the technologies, such as computers and telecommunications, within a constructivist context. Also, learning skills models and the main components of such courseware, so as to improve the resources available to the information society and enhance knowledge, are presented and discussed in [9]. How individual differences on cognitive styles, prior knowledge and gender influence the navigation pattern are elaborated by Somyurek et al. [10]. A methodology with conventional multimedia learning products involving the creation of interactive CDs is outlined by Barker et al. [11].

## 3 STRUCTURE OF CMC

The authors have considered a book as a tree structure where book title is the root of the tree. The first level children of the tree are the chapters of the book. The second level of the tree is the sections of the chapters. The third level is the sub-sections and the fourth level is the set of keywords that represents the sections or subsections. A courseware in general is based on multiple text books. Each text book has a unique tree structure. The tree structure for 'Database System Concept' by Korth et el. [12] is shown in Figure 1.

In the structures, the first level child node contains the chapter's title. Each chapter consists of a number of sections or subsections. So the second level nodes contain several titles of sections. Similarly the third level nodes contain a subsection title. The leaf nodes contain the keywords identifying the text of the sections or subsection.

In our representation, each subsection is represented by a unique path from root to keyword sets of the subsection. From Figure1, The authors observe that each text book has a hierarchical organization.

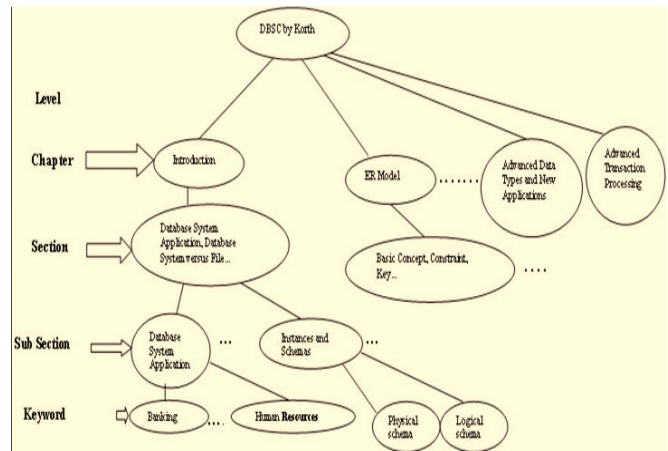

Figure 1: Tree structure for the book 'Database System Concept' by Korth et al.

If a courseware based on $n$ books, it has $n$ different hierarchical tree structures. The authors have to develop a common structure that covers all the topics of different books. It requires a dynamic merging or splitting of the structure. The general structure (Figure 2) independent of any book of the courseware can be formed by applying content clustering approach. Or a book structure can be considered as a standalone and a mapping can be found by applying the same approach.

## 4 GENERAL STRUCTURE OF GROUP COURSEWARE

The authors have considered a group of courses together and hence the root of the generalized tree (Figure 2) is the CMC group id. The reasons to consider a group courses are:

i. Multiple courses are usually developed based on similar multiple text books. As for example, Database Basic Course, Database Advanced Course, Distributed Database Course are based on similar text books like 'Database System Concept' by Korth et al. [12], 'Database management systems' by Ramakrishnan et al. [13] and 'Database Systems' by Davies et al. [14].

ii. Multiple text books are clustered together to help unified development of multiple similar courses.

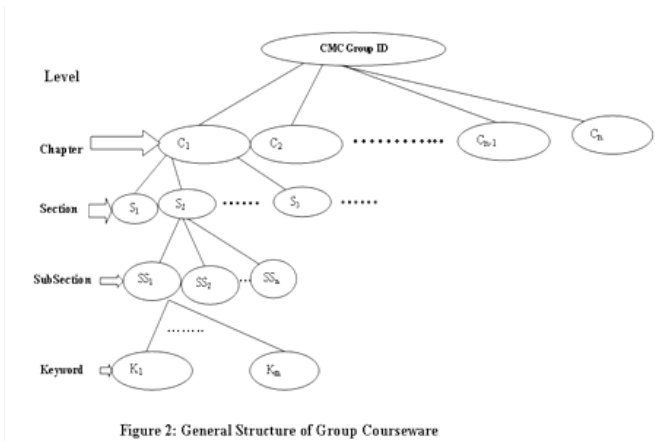

**Figure 2: General Structure of Group Courseware**

Now the group CMC can be modeled as follows:

Book = {chapters$_1$, chapters$_2$ … | any chapter$_i$ ∩ chapter$_j$= φ}

Chapters = {section$_1$, section$_2$ … | any section$_i$ ∩ section$_j$=φ}

Sections = {Subsection$_1$, Subsection$_2$ … | any Subsection$_i$ ∩ Subsection$_j$=φ}

Group Courseware = Book$_1$ ∪ Book$_2$ ∪…………∪ Book$_p$

$$= \{C_1, C_2, C_3 ………… C_n\}$$

Each $C_i$ is a cluster (equivalent to a chapter of a book) that is formed by applying clustering algorithm described in the following chapters.

In the generalized tree, an atomic content is represented by *CMC Group id*→$C_1$→$S_1$→$SS_1$→{$K_1$, $K_2$,……..,$K_m$ }. Now the authors can give a formal definition of our problem of clustering group courseware:

   i.   A group courseware consists of one or many books.

   ii.   Each book can be represented by a unique tree.

   iii.   All the books trees can be merged to form generalized tree structure.

> *Input:*
>     *B*: Set of books for group of courses;
>     *n*: number of books in the group.
> *Output:*
>     *GT*: Generalized tree covering B;
>     Algorithm GT Formation () {
>             *For* (i=1, i++, i<=n) {
>             **Construct BT (B$_i$)** for each book
> B$_i$;
>             GT= **Merge All BT ();**
>
>     Figure 3: Algorithm GT (Generalized tree) Formation.

Figure 3 represents the clustering of group courseware content by merging all individual book trees of a group courseware.

The merging is done by transforming all the book trees into a relational representation as described in the following section.

# 5 TRANSFORMATION OF BOOK TREE INTO RELATIONAL REPRESENTATION

To manage courseware tree the authors have represented each path of the tree as a tuple in the relational representation. As for example, the path (*DBSC by Korth*)Ⓘ*Introduction*Ⓘ(*Overview,……*) Ⓘ(*Database System Application*)Ⓘ(*banking, airlines,…..*) represents the first tuple in the relational representation for the book 'Database System Concept' by Korth et al. [12]. The schema relation is book-scheme (*Book Title, Chapter Title, Section Title, Subsection Title, Topic keyword*). For each subsection the authors have inserted one tuple in the relational representation. The authors have considered 'tuple id' as primary key which is useful for any kind of query.

# 6 CLUSTERING PROCEDURE

## 6.1 Selection of Keywords

The authors have considered the title text of the sections and subsection of a chapter to represent that chapter of the book. Considering only the topic keyword is not sufficient to represent a chapter, the proximity of the words has been considered as well. As for example, the individual consideration of 'database' and 'system' may mean different things than that of considering 'Database Management System'. A chapter is first identified by title of the chapter and then the keywords of the section/ subsection headings considering the proximity. After that, the authors have considered a few keywords to represent the section/ sub-section and topics within subsections. Firstly The authors have removed all stop words which are included with different level title for unification. As for example 'Entity Relation Model' is defined as a chapter by one author whereas 'The Entity Relation Model' by another. Moreover, the authors have converted all plural words to singular using grammatical rules and removed all hyphen or other characters which included in titles for matching.

Secondly the authors have set one book as reference clustering book and other books are used for mapping from the reference book. For this purpose, the authors have compared subsections keyword of different target books with different level keywords of a single reference book. As for example if a subsection be {A,B,C} keywords set of target book and {a,b,d}, {A,B,S}, {A,W,E,V} and {A,Q,S,W} be keywords set of different level of reference book then the authors have compared {A,B,C} with each level of reference book to find out matched keywords set for each level . To find out matched keywords set with {A,B,D} the authors have found the keyword set {A,B} in first iteration. In second iteration the authors have found

{A,B,AB}. Then final matched keywords set has been {A,B,AB} for chapter keyword. The authors have reduced C related all combinations. As a result comparisons have reduced effectively.

The authors have considered synonym table for matching. e.g., 'overview' is subsection keyword of target book and 'introduction' is chapter keyword of reference book by Korth et al [12]. In real fact introduction and overview are same thing. By synonym table the authors have found that these two are same. This similarity consideration has affected our relevance measurement positively. The authors have considered the authors ight for indicating relevance of a subsection with the chapter of reference book. The more the the authors ight, the more is the relevance. The authors have given highest relevance when a subsection of target book matches exactly with the chapter keyword of reference book. Because this matching has indicated proximity. The authors have reduced this relevance with top down approach of tree manner. That means when same keyword matches with down level keyword like section, subsection or topic keyword then it will be given less relevance.

When a matching keyword /set of keywords in full have not been matched with different levels of reference book then the authors have matched different combinations of the keyword. In those cases the authors have not considered proximity. Here also the authors have given different relevance value for different levels. Total relevance for a subsection of the target book with the chapter of reference book has been summed. By sorting and taking the highest relevance, the authors have found the chapter of reference book which is more relevant to the target subsection. In this way, the authors have found the total relevance table for target books of Ramakrisnan et al. [13] and Davies et al. [14] with reference book by Korth et al. [12]

## 6.2 Clustering Algorithm

The authors have applied following data mining clustering algorithm to form the required number of clusters as per requirement of the courseware.

```
Algorithm  SS_Clustering (ch_keyword, s_keyword, ss_keyword, t_keyword, ref_subsection)
    // ch_keyword, s_keyword, ss_keyword, t_keyword, are levels to represent a particular
    // chapter of target books. Sourcekeyword is  keyword representing Subsection keyword
    //of reference book, We ight is a global variable, rs1 represents record sets of reference book
    //rs2 represents record sets of target books
While rs1.EOF = False{
        Sourcekeyword= Call RemoveStopWord (ref_subsection)
        We ight=0;
        rs2.MoveFirst
    While rs2.EOF = False{
            // Measure with proximity
        Ch_Stop_Free = Call RemoveStopWord (ch_ keyword)
        Chap_MatchedKeySets = Call MatchingCount (Ch_Stop_Free, sourcekeyword)
        If (Ch_Stop_Free = Chap_MatchedKeySets) {
```

Figure 4: Algorithm for Subsection Clustering.

## 7 RESULT AND DISCUSSION

The authors have experimented our system with Pentium IV processor, 512 MB memory with 1.8 GHz Speed. The authors  have considered the database group of courseware based on three text books namely 1) 'Database System Concepts' by Korth et al. [12], 2) 'Database Management Systems' by Ramakrisnan et al. [13] and 3) 'Database Systems' by Davies et al. [14]. The number of chapters, sections and subsections for the above three books are shown in Table 1.

Table 1: Book information in detail

| Book Name | Authors Name | No. of Chapters | No. of Sections | No. of Subsections |
|---|---|---|---|---|
| Database System Concepts | Silberschatz, Korth, Sudarshan | 24 | 120 | 476 |

| Database Management Systems | Raghu Rama-krishnan, Johannes Gehrke | 28 | 191 | 373 |
|---|---|---|---|---|
| Database Systems | Paul Beynon-Davies | 43 | 172 | 382 |

The authors have considered 'Database Management Systems' by Ramakrisnan et al.[13] and 'Database Systems' by Davies et al. [14] as target books and 'Database System Concepts' by Korth et al. [12] as reference book. The authors have clustered 373 subsections of Ramakrishnan and 382 subsections of Davies with the reference book.

The authors have found out results for different no. of chapters of reference book for both target books. When the authors have clustered all subsections of target book with one chapter of reference book then the authors have found most of the subsections have come into false positive chapter of reference book. In that case the authors have also found no. of outliers is highest. An outlier is an observation that is numerically distant from the rest of the data. With the addition of more chapters of reference book the authors have found true positive increases exponentially and outlier decreases rapidly. Test result that accurately gives a positive reading is defined as true positive and negative reading is defined as false positive. When the authors have taken 24 chapters of the reference book by Korth et al. [12] the authors have found highest no. of true positive and lowest no. of false positive, outlier. The main reason of increasing true positive with more chapters is matching coverage spectrum increasing. Those subsections which have no relevance value by matching are considered as outlier. The authors have not considered subsections of lowest relevance as outlier because sometimes lowest relevance also may indicate true positive. Analyzing result the authors have got interesting character of true positive, false positive and outlier. Initially for less number of chapters false positive and outlier remains higher whereas true positive lower. This nature reverses with the increase of chapters. The Figure 5 and Figure 6 show it clearly.

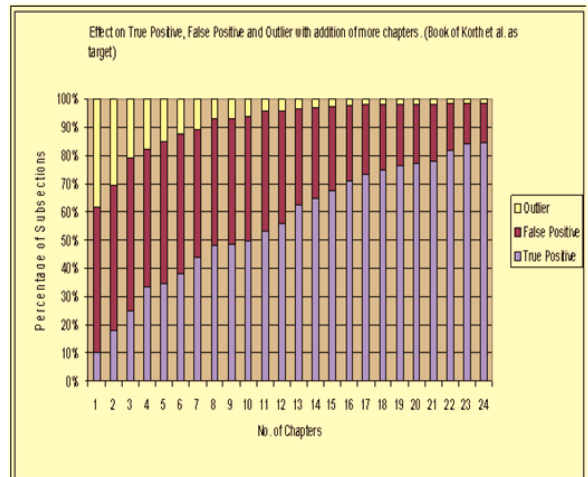

Figure 5: Effect on True Positive, False Positive and Outlier with addition of more chapters (Book of Korth et al. as target).

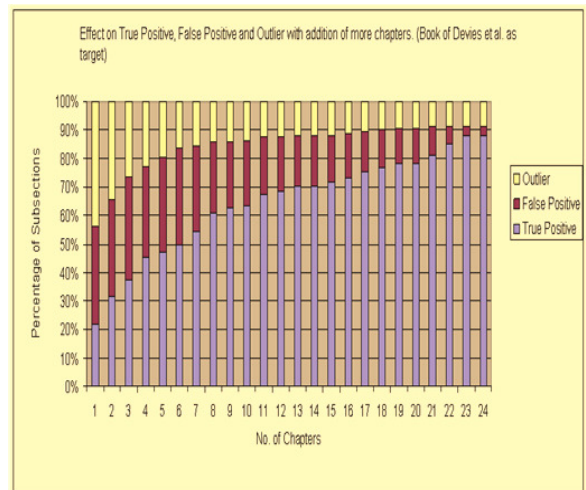

Figure 6: Effect on True Positive, False Positive and Outlier with addition of more chapters (Book of Devies et al. as target).

For both books the authors have to check false positive manually due to different organization, structure and style of books by different authors. Repetition is another reason for false positive. Homonym and synonym are also other problems for clustering. The authors have solved synonym problem effectively by using synonym table. The authors have also reduced the effect of homonym using correspondence of keyword set. Some false positives and outlier the authors have to handle manually due to these reasons. In this way the authors can take all books as target books and can be clustered to a single reference book. As a result a general structure that covers all topics of different books will be developed.

Though the authors have found some false positives in our result by domain expert but those are less than 15%. A very few number of outlier has occurred with both reference books when have been taken all chapters. For outlier indication the authors have not needed manual reck-oning. More than 85% of true positive have occurred when the authors have taken total spectrum of the reference book

## 7 CONCLUSION

Courseware so far has been developed for individual courses considering single book as a text and other books as mention. It is a necessity to cluster the contents of different books to form a generalized clustered content which is absent in the existing Computer Mediated Courseware. In this paper, a methodology has been proposed to construct a hierarchical general structure considering a group of courseware based on a set of text books. In that case, the authors have combined the individual structure of the set of books.

The authors have considered each book as a unique tree structure. Each text book has a hierarchical organization. The authors have transformed the trees into relational representation. Each unique path of the tree from root to leaves represents a tuple in that representation. By dynamic merging or splitting of the structure the authors have developed a general structure that covers all the topics of different books.

For clustering purpose the authors have considered proximity as well as synonym. The authors have considered one book as reference and the other as target. Each subsection of target book has been given a cumulative weight in accordance with matching with reference book for relevance. The more the weight, the more is the relevance. Similarity based clustering algorithm has been applied to find out the relevance. The authors have found clustering result for both target books taking the chapter which consists of highest relevance.

Result shows that very few numbers of subsections were found as outlier. The authors have assigned them to different chapters manually. The authors have verified our result by domain expert and found some false positive that were handled manually. The precision of our clustering algorithm has been found about 85%. If the authors could apply full text retrieval system, it would improve the performance sacrificing the computational time. This could be a future work.

## ACKNOWLEDGMENT

I want to express my cordial gratitude to my supervisor, Dr. Abu Sayed Md. Latiful Hoque, Associate Professor, Department of Computer Science and Engineering, Bangladesh University of Engineering and Technology, Dhaka. He gives me the opportunity to work in the field of database and make new idea in the database design strategies. The proper guideline of my supervisor helps me to complete this research. His extra ordinary knowledge on database helps me to find path in the vast area of database.

## REFERENCES

[1] en.wikipedia.org/wiki/Courseware
[2] R. Mason, "Models of online courses," ALN Magazine, 2, 2 (1998).
[3] http://www.cciencia.ufrj.br/educnet/eduead.htm
[4] http://www.uvex.edu/disted/definition.html
[5] J. Bourne, D. Harris, and F. Mayadas, "Online engineering education: learning anywhere, anytime," JALN, Vol. 9, Issue-1, March, 2005.
[6] A. J. Koppi, M. J. Chaloupka, and R. LleWellyn, "Computer mediated courseware development and the academic culture," In Proceedings of World Conference on Educational Telecommunications, Washington, June 19-24, 1999.
[7] V. P. Wade, and M. Lyng, "An automated evaluation service for educational courseware," In Proceedings of World Conference on the WWW and Internet, San Antonio, TX, October 30-November 04, 2000.
[8] N. Hoic-Bozic, J. Ledic, and J. Mezak, "Evaluating the use of World Wide We b courseware in student teachers' education: a case from Croatia," In Proceedings of Society for Information Technology & Teacher Education International Conference, San Diego, California, February 8-12, 2000.
[9] J. R. Galvao, and A. M. Barreto, "What is courseware? A comparative analysis," Vol. 4, Issue-2, World Transaction on Engineering and Technology Education, 2005.
[10] S. Somyurek, T. Guyer, and B. Atasoy, "The effects of individual differences on learner's navigation in a courseware," Vol. 7, Issue-2, The Turkish Online Journal of Educational Technology – TOJET, April, 2008.
[11] P. Barker, and S. Giller, "Models and methodologies for multimedia courseware production," In Proceedings of World Conference on Educational Multimedia, Hypermedia & Telecommunications, Denver, Colorado, June 24-29, 2002.
[12] A. Silberschatz, H. Korth, and S. Sudarshan, "Database System Concepts," ISBN: 9780071244763, AUG-05
[13] R. Ramakrishnan, and J. Gehrke, "Database management systems," 3RD Edition, ISBN: 10-0072465352, Jul 2007
[14] P. Beynon, and Davies, "Database Systems," 3RD Edition, ISBN: 1–4039–1601–2, 2004

**G. M. M. Bashir** received the B.Sc. Engg.degree in Computer Science and Engineering from Khulna University of Engineering and Technology, Khulna, Bangladesh, in 2003, and the M.Sc. in Information and Communication Technology from Bangladesh University of Engineering and Technology, Dhaka, Bangladesh, in 2009. Since 2006, he has been on the Department of Computer and Communication Engineering at Patuakhali Science and Technology University, Patuakhali, Bangladesh

**M. J. Hossain** has been on the Department of Computer Science and Information Technology at Patuakhali Science and Technology University, Patuakhali, Bangladesh

**M. R. Karim** received the B.Sc. Engg.degree in Computer Science and Engineering from Khulna University of Engineering and Technology, Khulna, Bangladesh, in 2003. He has been on Bangladesh Bank, Bangladesh